\def\RN{Reissner-Nordstr\"{o}m\ }
\def\rn{{\rm RN}}

\def\rfr#1{Eq. (\ref{#1})}

\def\derp#1#2{\rp{\partial{#1}}{\partial{#2}}}
\def\dert#1#2{\frac{{{d}}{#1}}{{{d}}{#2}}}              

\def\virg#1{``#1''}

\def\eqi{\begin{equation}}
\def\eqf{\end{equation}}
\def\eqia{\begin{eqnarray}}
\def\eqfa{\end{eqnarray}}
\def\rp#1#2{{#1\over#2}} \def\lb#1{\label{#1}}

\documentclass[11pt]{article}
\usepackage{graphicx}
\usepackage{dcolumn}
\usepackage{bm}
\usepackage{url}
\usepackage{amsmath,amsthm,amscd,amssymb}
\usepackage{latexsym,wasysym}
\usepackage{graphicx,epsfig}
\RequirePackage{color}

\begin{document}

\title{Constraining  the electric charges of some astronomical bodies  in  Reissner-Nordstr\"{o}m spacetimes and  generic $r^{-2}-$type power-law potentials from  orbital motions}

\author{L. Iorio \\ Ministero dell'Istruzione, dell'Universit\`{a} e della Ricerca (M.I.U.R.)-Istruzione \\ Fellow of the Royal Astronomical Society (F.R.A.S.) \\
 International Institute for Theoretical Physics and
Advanced Mathematics Einstein-Galilei \\ Permanent address: Viale Unit$\grave{\rm a}$ di Italia 68
70125 Bari (BA), Italy \\ email: lorenzo.iorio@libero.it}


\maketitle

\begin{abstract}
We  put \textcolor{black}{independent model dynamical} constraints on the net electric charge $Q$ of some astronomical and astrophysical objects by assuming that their exterior spacetimes are described by the  \RN\ metric, which induces an additional potential $U_{\rm RN}\propto Q^2 r^{-2}$. From the current bounds $\Delta\dot\varpi$ on any anomalies in the secular perihelion rate $\dot\varpi$ of Mercury and the Earth-mercury ranging $\Delta\rho$, we have $\left |Q_{\odot}\right|\lesssim 1-0.4\times 10^{18}\ {\rm C}$. Such constraints are $\sim 60-200$ times tighter than those recently inferred in literature. For the Earth, the perigee precession of the Moon, determined with the Lunar Laser Ranging (LLR) technique, and the  intersatellite ranging $\Delta\rho$ for the GRACE mission yield $\left|Q_{\oplus}\right| \lesssim 5-0.4\times 10^{14}\ {\rm C}$. The periastron rate of the double pulsar PSR J0737-3039A/B system allows to infer $\left|Q_{\rm NS}\right| \lesssim 5\times 10^{19}\ {\rm C}$. According to the perinigricon precession of the main sequence S2 star in Sgr A$^{\ast}$, the electric charge carried by the compact object hosted in the Galactic Center may be as large as $\left|Q_{\bullet}\right| \lesssim 4\times 10^{27}\ {\rm C}$. Our results extend to other hypothetical power-law interactions inducing extra-potentials $U_{\rm pert} = \Psi r^{-2}$ as well. It turns out that the \textcolor{black}{terrestrial GRACE mission}  yields the \textcolor{black}{tightest constraint} on the parameter $\Psi$, assumed as a universal constant, amounting to $\textcolor{black}{|\Psi|\lesssim 5\times 10^{9}\ {\rm m^4\ s^{-2}}}$.
\end{abstract}
\centerline

\section{Introduction}\lb{prima}
The \RN (RN hereafter) metric \cite{RN1,RN2,RN3} is the unique spherically symmetric and asymptotically flat solution of the Einstein-Maxwell equations.
It describes the exterior spacetime around an isolated spherical object of mass $M$ and electric charge $Q$. In standard isotropic coordinates \cite{RNiso}, commonly used for processing observations in several astronomical and astrophysical scenarios, the RN metric coefficients $g_{\mu\nu}$ are
\eqi
\begin{array}{lll}
g_{00} & = & 1+\rp{2 U}{c^2},\\ \\
g_{0i} & = & 0,\ i=1,2,3,\\ \\
g_{11} & = & g_{22}=g_{33}=-\left(1-\rp{2U}{c^2}\right),
\end{array}
\eqf
where $c$ is the speed of light in vacuum, and $U$ denotes the gravitational potential. For an isolated spherical body, it is
\eqi U = U_{\rm N}+U_\rn=-\rp{GM}{r}+\rp{k_e GQ^2 }{2 c^2 r^2},\eqf
where $Q$ is measured  in C,  $G=6.67384\times 10^{-11}\ {\rm N\ m^2\ kg^{-2}}$ is the Newtonian constant of gravitation, and
\eqi k_e \doteq \rp{1}{4\pi \varepsilon_0}=8.98755\times 10^{9}\ \rp{\rm N\ m^2}{\rm C^2},\eqf
in which $\varepsilon_0 =8.854187817\times 10^{-12}\ \rm{ N^{-1}\ m^{-2}\ C^2}$ is the electric constant.
Thus, the following constant factor can be defined
\eqi \xi_e\doteq \rp{k_e G}{2 c^2}=3.3363\times 10^{-18}\ {\rm  \rp{m^4}{C^2\ s^2}}.\lb{skala} \eqf

Concerning the physical relevance of $U_\rn$ in astronomical and astrophysical contexts, it relies upon the existence of macroscopic bodies stably endowed with net electric charges \cite{laslo}. Conversely, experimentally studying the orbital consequences of the \RN\ metric can allow, in principle, to measure, or, at least, constrain the value of the total net $Q$ carried by macroscopic bodies of astronomical size in a \textcolor{black}{dynamical independent model} way.

The question whether the Earth carries a net electric charge is quite old \cite{terra1,terra2}. Nonetheless, it is rather poorly defined \cite{terra3}. Indeed, it is often unclear if one refers to a quantity obtained by integrating over the solid and liquid Earth's surface-the \virg{globe}-, or to the whole \virg{planet} including the atmosphere and exosphere as well, with a deliberately set  outer boundary \cite{terra3}. Clearly, it is the second case that matters for our scopes, but it is difficult to have reliable estimates for it \cite{terra3}.  Concerning the Earth's surface, it should carry a net electrostatic negative charge of the order of \cite{terra2,terra4,terra5} $\left|Q^{\rm surf}_{\oplus}\right|\sim 4-5.7\times 10^5\ {\rm C}$ because the electrostatic potential near the surface of the Earth is about 100 V \cite{terra4,terra5}. According to the author of Ref. \cite{terra2}, tentative measurements of the net electric charge of the Earth as a planet would indicate that the electric field in the near-Earth space varies from  $\sim 0.1\ {\rm mV\ m}^{-1}$ to $\sim 10\ {\rm mV\ m}^{-1}$, implying $\left|Q_{\oplus}\right|\lesssim 50\ {\rm C}$.

As far as Sun-like, main sequence stars are concerned,  actually, they should not be exactly neutral, as recognized in the early twenties of the 20th century \cite{hale,Pann,Ross,Eddington,Cow,van}. Indeed,  the average velocity of free electrons in a plasma in thermodynamics equilibrium is higher than that of free protons, so that a larger number of electrons than protons on the stellar surface should attain the escape velocity, thus creating a net imbalance of positive charge \cite{Shva,bally,Neslu}. As a result, if $m_{\rm p}, m_{\rm e}, q_{\rm e}$  are the mass of the proton, the mass of  the electron and the charge of the electron, respectively, the expected total charge carried by a spherical, nonrotating star can be evaluated as \cite{Neslu}
\eqi Q_{\star}=\rp{2\pi\varepsilon_0 G(m_{\rm p}-m_{\rm e})}{q_{\rm e}}M_{\star}=77.03\ {\rm C}\eqf for $M_{\star}=\ {\rm M}_{\odot}$. Such a figure is in agreement with the bound set by the authors of Ref. \cite{bally} according to whom the net positive electric charge carried by a Sun-like, nonrelativistic star cannot exceed $\sim 100$ C. See also Ref. \cite{Glen}. On the other hand, in 1913 Hale \cite{hale}, in the framework of his attempts to detect the general magnetic field of the Sun, argued that the residual volume charge needed to account for it should have a negative sign.

Net electric charges in extreme astrophysical objects like white dwarfs, neutron stars and black holes may have profound influences on several fundamental processes occurring in such scenarios \cite{laslo2,Ruff}. E.g.,  an overcritical electric
field could occur in the region named dyadosphere \cite{dya} around \RN\ black holes, leading to the creation of $e^+-e^-$ pairs out of the vacuum which can cause  a \cite{cheru} Gamma Ray Burst (GRB) \cite{grb}.
The authors of Ref. \cite{Reise} deal with  nonrotating neutron stars in hydrostatic,
chemical (beta), and diffusive equilibrium. Baryons have high masses and are, therefore, kept inside
the star by its gravitational potential well, whose depth must be larger than the kinetic part of their Fermi energies. Anyway,
such a potential is not deep enough to hold the notably lighter electrons,
whose Fermi energy is relativistic (much larger than $m_{\rm e}c^2$) and which would therefore tend to escape, yielding a net
positive charge on the neutron star. However, at the typical densities occurring in neutron stars, already a
small charge imbalance between protons and electrons causes an
electrostatic field that can hold in the electrons and at the same
time partially balance the gravitational force on the protons, preventing
them from sinking into the center of the star, because of their relatively small kinetic energies.
Thus, such models of neutron stars contain a
spherically symmetric electrostatic potential evaluated as large as $U_{\rm e}\sim 10^8\ {\rm V}$ corresponding to a  charge of
\eqi Q_{\rm NS}\sim 111.26\ {\rm C}\eqf
by assuming $R_{\rm NS}=10\ {\rm km}$. The charge separation can occur also in rotating neutron stars endowed with  magnetic field and surrounded by plasma like pulsars \cite{Gold}. Although the model by the authors of Ref. \cite{Gold} presents shortcomings from the point of view of a realistic description \cite{Li}, it is nonetheless useful to illustrate some basic properties. Due to the complex interaction between the rotation of the star, its magnetic field and the surrounding plasma in its magnetosphere, a  $\rho_{\rm e}(r,\theta)$ charge density occurs, where $\theta$ is the colatitude. Assuming complete charge separation, the corresponding average number density is \cite{Lori}
\eqi \left\langle n_{\rm e}\right\rangle=1.75\times 10^{16}\ {\rm m^{-3}}\left(\rp{P}{\rm s}\right)^{-\rp{1}{2}}\left(\rp{\dot P}{10^{-15}}\right)^{\rp{1}{2}}, \lb{densita}\eqf
where $P$ is the rotation period of the neutron star. For \virg{normal} pulsars, i.e. for \cite{Lori} $P\sim 0.5\ {\rm s},\ \dot P\sim 10^{-15},$ \rfr{densita} yields ($r=R_{\rm NS}=10\ {\rm km}$)
\eqi Q_{\rm NS}\sim 1.7\times 10^{10}\ {\rm C},\eqf
while for millisecond pulsars with \cite{Lori} $P\sim 3\ {\rm ms},\ \dot P\sim 10^{-20}$ it is
\eqi Q_{\rm NS}\sim 6.8\times 10^{8}\ {\rm C}.\eqf
For highly compact stars, whose radius is on the verge of forming an event
horizon, the extremely high density and relativistic effects may, in principle,  affect also the net charge of
a compact star, so that it can take much more charge to be in equilibrium \cite{cold1,cold2,cold3,cold4,cold5,cold6}. The authors of Ref. \cite{cold7}
pursue the ideas of the aforementioned works by dealing with the effect of
charge in cold compact stars, made of neutrons, protons and
electrons They assume that the charge density is proportional to
mass density, and that the net charge in the system is in the form
of trapped charged particles carrying positive electric charge. By posing $Q_{\rm NS}\sim \sqrt{G 4\pi\varepsilon_0}M$, the authors of Ref. \cite{cold7} find
$Q_{\rm NS}\sim 2\times 10^{20}\ {\rm C}$. It turns out \cite{cold7} that such highly electrified stars would not be stable, collapsing to a black hole: during the collapse, very little amount of charge would have the opportunity to escape, so that a charged black hole would form.
Lensing by a charged neutron star, whose exterior spacetime is modeled by the \RN\  metric, was studied in Refs. \cite{Dab1,Dab2}.

The authors of Ref. \cite{quasi} investigated charged frozen stars. They are quasiblack holes with pressure, i.e. objects whose boundary approaches their
own gravitational radius as closely as one likes, without actually reaching it because of strong internal relativistic effects. Such spherically symmetric relativistic charged fluid distributions are bounded by a physical surface of radius $R_0$. Their internal region is filled with a fluid characterized by its mass and charge densities and by a nonzero pressure. The exterior spacetime  is described  by the \RN\  metric  \cite{quasi}.

Electrically charged strange quark stars were studied in Ref. \cite{strange}. Such objects would be compact stars made of absolutely stable strange quark matter. They may carry ultrastrong electric fields on their surfaces, of the order of $\sim 10^{20}\ {\rm V\ m^{-1}}$ for ordinary strange matter. Under certain circumstances, the strength of the electric field may increase to values that exceed $10^{21}\ {\rm V\ m^{-1}}$.

For electric charges in black holes, see \cite{Wald74,Wald,Heusle,Puns}.
The formation of \RN\ black holes was intensively studied in recent years, also because the astrophysical
conditions  leading to it may look rather problematic: see, e.g., Refs. \cite{forma1,forma2,forma3,forma4,forma5,cold7,forma6,forma7,forma8,forma9,forma10,forma11,forma12}.
Anyway,  the existence of such a kind of black holes is neither forbidden by theoretical nor observational
arguments \cite{Zak}. Also the possibility that electrically charged rotating black holes, described by the Kerr-Newman (KN) spacetime metric \cite{KN1,KN2}, really exist in nature  is somewhat controversial \cite{Puns2}.
Gravitational lensing in \RN\ spacetimes was studied in Ref. \cite{lensing1} on the basis of the results of Refs. \cite{lensing2,lensing3,lensing4,lensing5}.
For some features of Schwarzschild black hole lensing, see, e.g., Ref. \cite{virbha0a,virbha0b}.
Methods to measure the electric charge of black holes through retrolensing with the currently ongoing \cite{launch} astrometric mission Spektr-R (known also as Radioastron) \cite{radioastron} were proposed in Ref. \cite{Zak}; see also Refs. \cite{Zak2,depa}.
There is nowadays wide consensus that the compact object with $M_{\bullet}=4\times 10^6\ {\rm M}_{\odot}$  hosted by the Galactic Center in Sgr A$^{\ast}$ \cite{ghez} is a supermassive black hole. It is orbited by several main sequence S stars: the closest one so far discovered, known as S2, revolves around Sgr A$^{\ast}$ at $\left\langle r \right\rangle = 1433$ AU from it in $15.98$ yr \cite{gillessen}. The author of Refs. \cite{binun1,binun2} pointed out that lensed images of the stars orbiting close to Sgr A$^{\ast}$ can provide insight into the form of its exterior spacetime metric and, in particular, on its electric charge $Q_{\bullet}$ as well. Gravitational lensing of stars orbiting the supermassive black hole in the Galactic Center was studied  in Ref. \cite{bozza} as well. According to the author of Ref. \cite{Zak2}, for Sgr A$^{\ast}$, a charged black hole
could be distinguished from a Schwarzschild black hole with the space radio telescope Radioastron, at least if its charge is close to the extremal value
\eqi Q^{\rm extr}_{\bullet}=\sqrt{G 4\pi\varepsilon_0} M_{\bullet}=7.4\times 10^{26}\ {\rm C}\lb{limiteq}\eqf
allowed to avoid a naked singularity. In general, black holes and naked singularities can be observationally differentiated  through their gravitational lensing characteristics \cite{virbha1,virbha2,virbha3}. We recall that there are no logical arguments preventing existence of naked singularities; according to Ref. \cite{penrose}, this is an open question.  In the case of a Kerr-Newman metric, $Q^{\rm extr}_{\bullet}$ is modified by the black hole's spin $J_{\bullet}$ as
\eqi Q^{\rm extr}_{\bullet}=\sqrt{G 4\pi\varepsilon_0\left(1-\chi_{\bullet}^2\right)} M_{\bullet}=(6.3-6.6)\times 10^{26}\ {\rm C}\eqf
since for Sgr A$^{\ast}$ it is \cite{Kato,Ghenz}
\eqi \chi_{\bullet}\doteq\rp{J_{\bullet}c}{M_{\bullet}^2 G} \sim 0.44-0.52.\eqf
For a \RN\  black hole, its charge changes the size of the shadows up to $30\%$ in the extreme charge case. Therefore, the charge of the black hole can be measured by observing the shadow size, if the other black hole parameters are known with sufficient precision \cite{Zak2}.
The emission and absorption properties of charged black holes were studied in Refs. \cite{emission1,emission2,emission3,emission4}. Their characteristic pattern could be used as footprint of the black hole. For quantum mechanical uncertainties in measuring, among other things,
the charge  of a black hole, see Ref. \cite{Oha}.
The geodesics  in the \RN\ spacetime were studied in Ref. \cite{Chandras}.

The appearance of charged, traversable wormholes in the \RN\ metric is investigated by the authors of Ref. \cite{worm}.

In Sec. \ref{seconda} of this paper we will use well known and largely tested orbital motions (see Sec. \ref{perielio}) of some natural (Sec. \ref{pianeti}) and artificial (Sec. \ref{grazia})  bodies in the solar system  to infer constraints on the net electric charges $Q$ carried by the Sun and the Earth.
In Sec. \ref{neutrona} we will also use  the well known, and extensively studied, double pulsar binary system and the main sequence  S2 star orbiting the compact object in Sgr A$^{\ast}$. The resulting constraints are inferred in a phenomenological, model-independent way; the only assumption made is that the exterior spacetime of the sources of the gravitational field is described by the \RN\ metric. For the motion of electrically and magnetically charged particles in such a spacetime, see Refs. \cite{Pugl,Grun}.
Finally, we stress that our results are not necessarily limited to the \RN\ spacetime, being valid also for other theoretical schemes yielding  power-law interaction potentials $U_{\rm pert}\propto r^{-2}$ \cite{adelberger}. This point is treated in Sec. \ref{extrapot}.

\section{\RN\ long-term orbital effects and comparison with the observations}\lb{seconda}
\subsection{Analytical calculation of the secular precession of the pericenter}\lb{perielio}
The long-period effects caused by $U_\rn$ on the motion of an electrically neutral particle of mass $m$ can, e.g.,  be
computed perturbatively by adopting the Lagrange equations for the variation of the osculating Keplerian \textcolor{black}{orbital} elements \cite{capderou}: their validity has been confirmed in a variety of independent phenomena.
Generally speaking, the \textcolor{black}{Lagrange equations} imply the use of a perturbing function $\mathcal{R}$ which is the correction $U_{\rm pert}$ to the standard Newtonian monopole \textcolor{black}{$U_{\rm N}\propto r^{-1}$}.
In the case $U_{\rm pert}=U_\rn$, the average over one orbital revolution of the perturbing function $\mathcal{R}$ is straightforwardly obtained by using the true anomaly $f$ as fast variable of integration. It is
\eqi \left\langle\mathcal{R}\right\rangle = \rp{k_eGQ^2}{ 2c^2a^2\left(1-e^2\right)^{\rp{1}{2}}},\lb{pertfun}\eqf
where $a$ is the semimajor axis and $e$ is the eccentricity of the test particle's orbit.
From \rfr{pertfun} and  the Lagrange equation for variation of the longitude of pericenter $\varpi$ \cite{bertotti}
\eqi \left\langle\dert\varpi t\right\rangle = -\rp{1}{n_{\rm b} a^2}\left\{\left[\rp{\left(1-e^2\right)^{\rp{1}{2}}}{e}\right]\derp{\left\langle\mathcal{R}\right\rangle}e + \rp{\tan\left(\rp{I}{2}\right)}{\left(1-e^2\right)^{\rp{1}{2}}}\derp{\left\langle\mathcal{R}\right\rangle}I\right\}, \lb{lagra}\eqf
it turns out that $\varpi$ experiences a secular precession given by
\eqi \left\langle\dert\varpi t\right\rangle = -\rp{k_e GQ^2}{2 c^2 n_{\rm b} a^4 \left(1-e^2\right)}=-\rp{k_e G^{\rp{1}{2}}Q^2}{2c^2 M^{\rp{1}{2}} a^{\rp{5}{2}}\left(1-e^2\right)}.\lb{precess}\eqf
In \rfr{lagra}\textcolor{black}{-\rfr{precess}}, $n_{\rm b}\doteq\sqrt{GM/a^3}$ is the Keplerian mean motion; $I$, \textcolor{black}{entering \rfr{lagra},} is the inclination of the orbital plane to the reference   $\{x,y\}$ plane. The longitude of pericenter $\varpi\doteq\Omega+\omega$ is a \virg{dogleg} angle since it is the sum of the longitude of the ascending node $\Omega$, which is an angle in the reference $\{x,y\}$ plane from a reference $x$ direction to the intersection of the orbital plane with the $\{x,y\}$  plane itself (the line of the nodes), and of the argument of pericenter $\omega$, which is an angle counted in the orbital plane from the line of the nodes to the point of closest approach, usually dubbed  pericenter. The precession of \rfr{precess}, which is an exact result in the sense that the \textcolor{black}{assumption $e\sim 0$ was not} made, agrees with the one obtained by Adkins and McDonnel in Ref. \cite{adkins} with a more cumbersome calculation. To facilitate a comparison between such two results, we note that, in general, the authors of Ref. \cite{adkins} work out the pericenter advance per orbit $\Delta\theta_p$: it corresponds to our $\Delta\varpi=\left\langle\dot\varpi\right\rangle {\rm P}_{\rm b},$ where ${\rm P}_{\rm b}\doteq 2\pi/n_{\rm b}$ is the orbital period. Moreover, in the potential energy $V(r)=\alpha_{-(j+1)}r^{-(j+1)}$ of Ref. \cite{adkins} it must be posed $j=1$ and $\alpha_{-2}\rightarrow 2^{-1} c^{-2} k_e G m Q^2$, while in $\Delta\theta_p(-(j+1))$ of Eq. (38) in Ref. \cite{adkins} it must be set $L\rightarrow a(1-e^2),\ \chi_1(e)=2$. With such replacements, it can be shown that the advance per orbit of Eq. (38) in Ref. \cite{adkins} corresponds just to our precession in \rfr{precess}. The precession of \rfr{precess} yields a pericenter advance per orbit $\Delta\varpi$ which, by posing  $L\rightarrow a(1-e^2)$,  agrees also with the  shift calculated by the authors of Ref. \cite{avalos} in their Eq. (16) from the geodesic equations of motion, and in their Eq. (37) with the Laplace-Runge-Lenz vector \cite{LRL1,LRL2}.
The rate of \rfr{precess} also agrees with that obtainable from the fractional shift computed by Lemmon and Mondragon \cite{Lemm} as $(2\pi)^{-1}\Delta\varphi=-\epsilon_Q$, where $\epsilon_Q\doteq 2\epsilon_M r_Q^2 r_M^{-2},\ r_M\doteq 2GM c^{-2},\ r_Q^2\doteq k_e Q^2 G c^{-4},\ \epsilon_M\doteq  GMc^{-2}a^{-1}(1-e^2)^{-1}$. For earlier derivations of the pericenter precession in the \RN\ metric,  see Refs. \cite{Jaffe,Teli,Wanas,Chal}.

The analytical result of \rfr{precess} is confirmed by a numerical integration of the equations of motion for Mercury with, say, $Q_{\odot}=10^{19}\ {\rm C}$   displayed in Figure \ref{figura1}: both yield $-24$ milliarcseconds per century (mas cty$^{-1}$ in the following). The choice of the numerical value adopted for the electric charge of the Sun is purely arbitrary, being motivated only by the need of dealing with easily manageable figures on the vertical axis.
\begin{figure}
\begin{center}
\epsfig{file=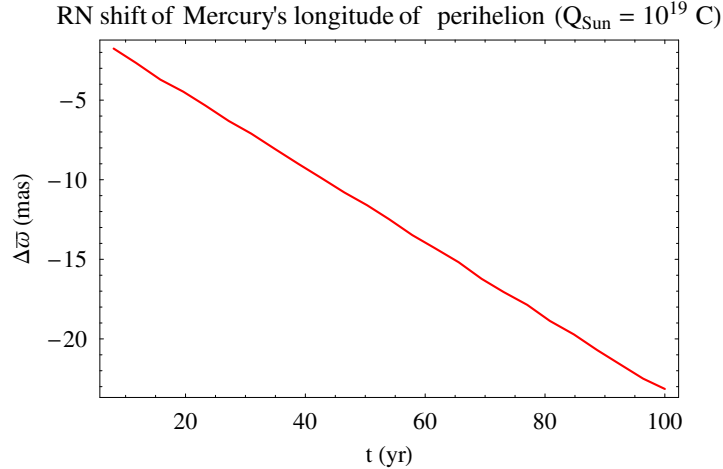}
\caption{Centennial shift $\Delta\varpi$, in milliarcseconds (mas), of the longitude of perihelion of Mercury due to the \RN\  potential of the Sun for, say, $Q_{\odot}=10^{19}\ {\rm C}$, chosen just for illustrative purposes,  computed from the difference between  two numerical integrations of the equations of motion in cartesian coordinates  performed with and without $U_{\rm RN}$. Both the integrations share the same initial conditions, retrieved from the NASA WEB interface HORIZONS at http://ssd.jpl.nasa.gov/horizons.cgi, for the epoch J2000.0. It agrees with the analytical result of \rfr{precess}. }\lb{figura1}
\end{center}
\end{figure}

The authors of Ref. \cite{Kezer} computed the correction ${\rm P}_Q$ to the Keplerian orbital period ${\rm P}_{\rm b}$ due to the \RN\ metric in the framework of the studies about general relativistic effects on  bound orbits of solar sails \cite{sail1,sail2}.
\subsection{Constraints on $Q_{\odot}$ from solar system planetary orbital motions}\lb{pianeti}
The corrections $\Delta\dot\varpi$ to the standard Newtonian-Einsteinian secular precessions of the longitudes of the perihelia are routinely used by independent teams of astronomers \cite{pitjeva,fienga} as a quantitative measure of the maximum size of any putative anomalous effect allowed by the currently adopted mathematical models of the standard solar system dynamics fitted to the available planetary observations. Thus, $\Delta\dot\varpi$ can be used to put constraints on the parameters like $Q$ entering the  models one is interested in. From \rfr{precess} it turns out that the tightest constraints come from Mercury, which is the innermost planet with $a=0.38$ AU.

Fienga et al. \cite{fienga}, who used also a few data from the three flybys of MESSENGER in 2008-2009, released an uncertainty of $0.6$ mas cty$^{-1}$  for the perihelion precession of Mercury, so that it is \eqi Q_{\odot}\lesssim 1.5\times 10^{18}\ {\rm C}.\eqf The uncertainty in the pre-MESSENGER Mercury's perihelion extra-precession by Pitjeva \cite{pitjeva} is about one order of magnitude larger (5 mas cty$^{-1}$).

Tighter constraints on $Q_{\odot}$ come from the interplanetary Earth-Mercury ranging. Indeed, according to Table 1 of Ref. \cite{fienga}, the standard deviation $\sigma_{\Delta\rho}$ of the Mercury range residuals $\Delta\rho$, obtained with the INPOP10a ephemerides and including also three Mercury MESSENGER flybys in 2008-2009, is as large as $1.9$ m. A numerical integration of the equations of motion of Mercury and the Earth with \eqi Q_{\odot}\lesssim 4.2\times 10^{17}\ {\rm C}\eqf  yields a \RN\  range signal with the same standard deviation; Figure \ref{figura2} displays it.
\begin{figure}
\begin{center}
\epsfig{file=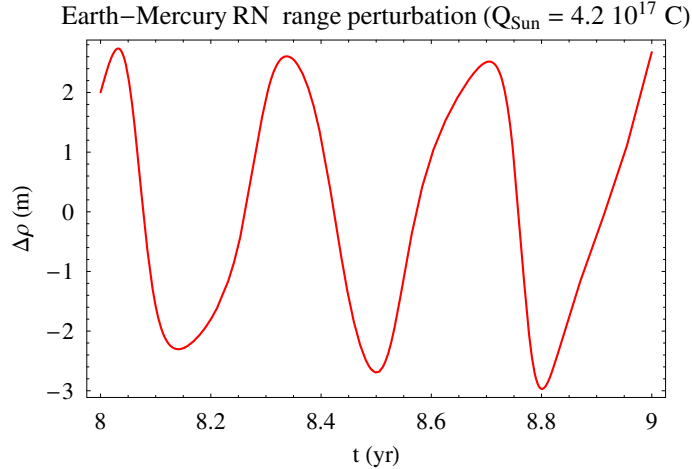}
\caption{2008-2009 Earth-Mercury range shift $\Delta\rho$, in m, due to the unmodeled \RN\ potential of the Sun for $Q_{\odot}=4.2\times 10^{17}\ {\rm C}$ computed from the difference between  two numerical integrations of the equations of motion in cartesian coordinates  performed with and without $U_{\rm RN}$. Both the integrations share the same initial conditions retrieved from the NASA WEB interface HORIZONS at http://ssd.jpl.nasa.gov/horizons.cgi, for the epoch J2000.0. It is $\left\langle \Delta\rho \right\rangle=0.04\ {\rm m},\ \sigma_{\Delta\rho}=1.9\ {\rm m}$. }\lb{figura2}
\end{center}
\end{figure}

Our model-independent, dynamical bounds for the electric charge of the Sun are $\sim 60-200$ times more stringent than $Q_{\odot}\lesssim 8.91\times 10^{19}\ {\rm C}$ inferred by the authors of Ref. \cite{avalos}.
\subsection{Constraints on $Q_{\oplus}$ from the Moon and the GRACE spacecraft orbiting the Earth}\lb{grazia}
Remaining within the solar system, let us, now, constrain the electric charge of the Earth with natural and artificial bodies.

 The orbit of the Moon is accurately reconstructed with the Lunar Laser Ranging (LLR) technique \cite{LLR} since 1969; Figure B-1 of Ref. \cite{folkner} shows that the residuals
of the Earth-Moon range are at a cm-level since about 1990. The secular precession of the lunar perigee is known with an accuracy of about $0.1\ {\rm mas\ yr}^{-1}$ \cite{accura1,accura2,accura3,accura4}; thus, \rfr{precess} yields \eqi Q_{\oplus}\lesssim 5.1\times 10^{14}\ {\rm C}.\eqf

The Gravity Recovery and Climate Experiment (GRACE) mission \cite{grace}, jointly launched in March 2002 by NASA and the German Space Agency (DLR) to  map the terrestrial gravitational field with an unprecedented accuracy, consists of a tandem of two spacecrafts moving along low-altitude, nearly polar orbits  continuously linked by a Satellite to Satellite Tracking (SST) microwave K-band ranging (KBR) system accurate to better than $10\ \mu$m (biased range $\rho$) \cite{range} and $1\ \mu$m s$^{-1}$ (range-rate $\dot\rho$) \cite{range,rrate}. Studies for a follow-on of GRACE \cite{follow} show that the use of a  interferometric laser ranging system may push the accuracy in the range-rate to a $\sim 0.6\ {\rm nm\ s^{-1}}$ level. A numerical integration of the equations of motion for GRACE A/B, including also the mismodelled signal of the first nine zonal harmonics of geopotential \cite{capderou} according to the global Earth gravity model GOCO01S \cite{goco01s}, shows that the SST range is more effective than the SST range-rate in constraining the Earth's electric charge  for which it holds \eqi Q_{\oplus}\lesssim 4\times 10^{13}\ {\rm C},\lb{graceQ}\eqf which is about one order of magnitude better than the lunar constraint. Figure \ref{figura3} depicts the numerically integrated GRACE SST range signal due to $U_{\rm RN}$ and the aforementioned mismodeled zonals.
\begin{figure}
\begin{center}
\epsfig{file=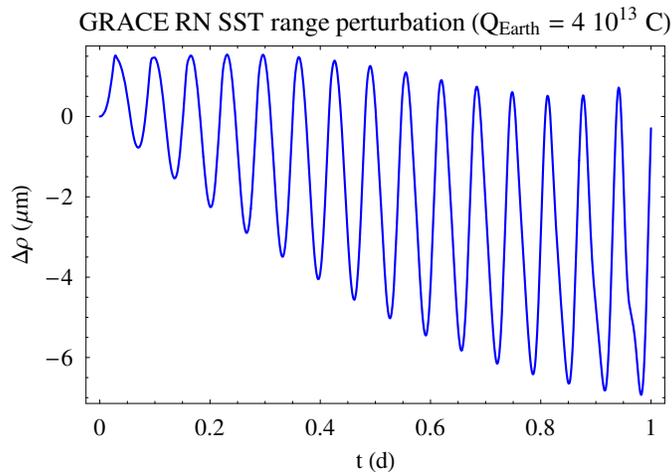}
\caption{Daily GRACE SST range shift $\Delta\rho$, in $\mu$m, due to the unmodeled \RN\ potential of the Earth ($Q_{\oplus}= 4 \times 10^{13}\ {\rm C}$) and the first nine mismodelled zonals of geopotential according to GOCO01S \cite{goco01s} computed from the difference between  two numerical integrations of the equations of motion in cartesian coordinates  performed with and without $U_{\rm RN}$. Both the integrations share the same initial conditions contained in the files GNV1B$\_$2003-09-14$\_$A$\_$00 and  GNV1B$\_$2003-09-14$\_$B$\_$00 retrieved from ftp://cddis.gsfc.nasa.gov/pub/slr/predicts/current/graceA$\_$irvs$\_$081202$\_$0.gfz and ftp://cddis.gsfc.nasa.gov/pub/slr/predicts/current/graceB$\_$irvs$\_$081201$\_$1.gfz. The epoch is 13 September 2003. (See ftp://podaac.jpl.nasa.gov/pub/grace/doc/Handbook$\_$1B$\_$v1.3.pdf for the explanation of the GPS Navigation Data Format Record (GNV1B) format).
It is $\left\langle \Delta\rho \right\rangle=-2\ \mu{\rm m},\ \sigma_{\Delta\rho}=2\ \mu{\rm m}$. }\lb{figura3}
\end{center}
\end{figure}
It may be interesting to notice that the GRACE-based dynamically inferred upper bound of \rfr{graceQ} almost corresponds to the net electric charge ($Q_{\oplus}=2.9\times 10^{13}\ {\rm C}$, from $U_{\rm e} = 4.1\times 10^{16}$ V) that the Earth's surface should have carried to account for the terrestrial magnetic field in the early theory propounded in 1879 by Perry and Ayrton  \cite{perry}. It was soon rejected by Rowland \cite{rowland} because of the manifest lacking of the enormous effects implied by it.
\subsection{Constraints from the double pulsar and Sgr A$^{\ast}$}\lb{neutrona}
The periastron  of the double pulsar PSR J0737-3039A/B  \cite{burgay,lyne} can be used to constrain $Q_{\rm NS}$, at least for this particular system.  The present-day accuracy in measuring the secular precession of the periastron is $6.8\times 10^{-4}$ degree per year (deg yr$^{-1}$ in the following) \cite{kramer}; thus, a straightforward application of it to \rfr{precess} would give \eqi Q_{\rm NS}\lesssim 6.9\times 10^{18}\ {\rm C}.\eqf Actually, the larger uncertainty in the theoretical expression of the general relativistic 1PN periastron precession must be taken into account as well. It is as large as $0.03$ deg yr$^{-1}$ \cite{iorio}, so that it yields \eqi Q_{\rm NS}\lesssim 4.7\times 10^{19}\ {\rm C}.\lb{limiteqns}\eqf Notice that the constraint of \rfr{limiteqns} is smaller by almost one order of magnitude than $\sqrt{G 4\pi\varepsilon_0} M_{\rm tot}= 4.4\times 10^{20}\ {\rm C}$ for the pulsar PSR J0737-3039A/B system. \textcolor{black}{For other electromagnetic/gravitational effects occurring in binary systems hosting compact objects, see, e.g., Ref. \cite{capozz2} }.

The perinigricon of the S2 star, orbiting in $15.98$ yr the compact object hosted in Sgr A$^{\ast}$ with $M=4\times 10^6 M_{\odot}$ \cite{ghez} at $\left\langle r\right\rangle = 1433$ AU from it \cite{gillessen} can be used to put dynamical constraints on the electric charge of the attractive center. Indeed, an accuracy of \cite{gillessen}
$0.05\ {\rm deg\ yr}^{-1}$ can be inferred for the perinigricon precession of S2: thus, \eqi Q_{\bullet}\lesssim 3.6\times 10^{27}\ {\rm C}.\lb{qBH}\eqf By propagating the uncertainties in the estimated parameters of the Sgr A$^{\ast}$-S2 system \cite{gillessen} entering \rfr{precess}, it is possible to infer \eqi\sigma_{Q_{\bullet}}=4.6\times 10^{26}\ {\rm C}.\lb{dqBH}\eqf In view of \rfr{limiteq}, this implies that the present-day knowledge of the S2 orbital motion still leaves room for the possibility that the massive object in the Galactic Center is not a \RN\ black hole since its electric charge might be larger than the extremal value to avoid a naked singularity. Discovering and monitoring other stars, closer to Sgr A$^{\ast}$, will be crucial to shed further light on such an issue. \textcolor{black}{For other potentially competing orbital effects which may take place in dense stellar cluster at the Galactic Center, see. Ref. \cite{capozz}.}
\section{Constraints on a generic $r^{-2}$ extra-potential}\lb{extrapot}
The bounds previously obtained in terms of electric charges in the \RN\ spacetime can also be used to constrain the parameter $\Psi$ of a generic  $r^{-2}$ extra-potential \cite{adelberger}
\eqi U_{\rm pert}=\rp{\Psi}{r^2},\eqf provided that the replacement
\eqi \xi_e Q^2\rightarrow \Psi\eqf is made: $[\Psi]={\rm L^4\ T^{-2}}$.
Here we will adopt a purely phenomenological approach by considering $\Psi$ just as an universal constant independent of the body which acts as source of the (modified) potential; for a discussion of power-law potentials, see Ref. \cite{adelberger}. In a quantum field theory framework, such kind of interactions arise from higher-order exchange processes with simultaneous exchange of multiple massless bosons. In particular, potentials $\propto r^{-2}$ are generated by the simultaneous exchange of two massless scalar bosons \cite{boh}.

It turns out that \textcolor{black}{GRACE} yields the tightest constrain on $\Psi$, considered as a universal parameter, amounting to
\eqi\textcolor{black}{ \left|\Psi\right|\lesssim 5.3\times 10^{9}\ {\rm m^4\ s^{-2}}}.\eqf
Among the solar planets, the most stringent bound comes from the \textcolor{black}{Earth-Mercury ranging} with
\eqi \textcolor{black}{\left|\Psi\right|\lesssim 6\times 10^{17}\ {\rm m^4\ s^{-2}}. }\eqf The double pulsar and the Sgr A$^{\ast}$-S2 system yields $\textcolor{black}{\left|\Psi\right|\lesssim 7.4\times 10^{21}\ {\rm m^4\ s^{-2}}}$ and $\left|\Psi\right|\lesssim 4.3\times 10^{37}\ {\rm m^4\ s^{-2}}$, respectively.
\textcolor{black}{
\section{Summary and conclusions}
We analytically worked out the long-term precession of the pericenter of a test particle orbiting a central body of mass $M$ and electric charge $Q$ in the framework of the Reissner-Nordstr\"{o}m metric. We obtained a non-vanishing, secular precession whose expression is valid for any value of the eccentricity $e$ of the particle's orbit.}

\textcolor{black}{Then, we compared our analytical result to latest observations in several astronomical and astrophysical scenarios to infer upper bounds on the electric charge of various objects acting as sources of the Reissner-Nordstr\"{o}m field. As far as our solar system is concerned,  we get $Q_{\odot}\lesssim 4\times 10^{17}$ C from Mercury's motion. The GRACE mission around the Earth yields $Q_{\oplus}\lesssim 4\times 10^{13}$ C. Planned follow-on of GRACE and spacecraft-based missions to Mercury like MESSENGER (ongoing) and BepiColombo (to be launched in next years) will allow to strengthen such constraints. The periastron advance of the double pulsar PSR J0737-3039A/B, corrected for the uncertainty in the standard 1PN precession, yields $Q_{\rm PSR}\lesssim 5\times 10^{19}$ C. The orbital motion of the main sequence S2 star around the supermassive black hole in Sgr A$^{\ast}$, monitored during a time span larger than a full orbital period, allows to infer $Q_{\bullet}\lesssim 4\times 10^{27}$ C. This bound could become tighter in future if stars orbiting Sgr A$^{\ast}$ faster than S2 will be discovered and monitored.}

\textcolor{black}{Our results are not necessarily limited to the Reissner-Norstr\"{o}m spacetime due to a charged body, being, instead, valid for any perturbing potential $U_{\rm pert}\propto r^{-2}$ through a universal parameter $\Psi$ having dimensions of $[\Psi]={\rm L^4\ T^{-2}}$. In this respect, the tightest constraints come from the Earth-GRACE system with $|\Psi|\lesssim 5\times 10^9$ m$^4$ s$^{-2}$.
}
\section*{Acknowledgements}
I thank K.  S.  Virbhadra for stimulating correspondence.


\end{document}